\documentclass[twocolumn]{aastex631}

\providecommand{\bjdtdb}{\ensuremath{\rm {BJD_{TDB}}}}
\providecommand{\feh}{\ensuremath{\left[{\rm Fe}/{\rm H}\right]}}
\providecommand{\teff}{\ensuremath{T_{\rm eff}}}

\providecommand{\msun}{\ensuremath{\,M_\Sun}}
\providecommand{\rsun}{\ensuremath{\,R_\Sun}}

\providecommand{\mj}{\ensuremath{\,M_{\rm J}}}
\providecommand{\rj}{\ensuremath{\,R_{\rm J}}}

\providecommand{\logg}{$\log{g}$}
\providecommand{\kms}{km~s$^{-1}$}

\usepackage{appendix}

\usepackage{hyperref}

\begin{document}

\title{Searching for GEMS: Discovery and Characterization of Two Brown Dwarfs Around M Dwarfs\footnote{Based on observations obtained with the Hobby-Eberly Telescope (HET), which is a joint project of the University of Texas at Austin, the Pennsylvania State University, Ludwig-Maximillians-Universitaet Muenchen, and Georg-August Universitaet Gottingen. The HET is named in honor of its principal benefactors, William P. Hobby and Robert E. Eberly.}}

\author[0000-0002-0786-7307]{Alexander Larsen}
\affiliation{Department of Physics \& Astronomy, University of Wyoming, Laramie, WY 82070, USA}

\author[0000-0002-5817-202X]{Tera N. Swaby}
\affiliation{Department of Physics \& Astronomy, University of Wyoming, Laramie, WY 82070, USA}

\author[0000-0002-4475-4176]{Henry A. Kobulnicky}
\affiliation{Department of Physics \& Astronomy, University of Wyoming, Laramie, WY 82070, USA}

\author[0000-0003-4835-0619]{Caleb I. Ca\~nas}
\affiliation{NASA Goddard Space Flight Center, 8800 Greenbelt Road, Greenbelt, MD 20771, USA}

\author[0000-0001-8401-4300]{Shubham Kanodia}
\affiliation{Earth and Planets Laboratory, Carnegie Institution for Science, 5241 Broad Branch Road, NW, Washington, DC 20015, USA}

\author[0000-0002-2990-7613]{Jessica Libby-Roberts}
\affiliation{Department of Astronomy \& Astrophysics, The Pennsylvania State University, 525 Davey Laboratory, University Park, PA 16802, USA}
\affiliation{Center for Exoplanets and Habitable Worlds, The Pennsylvania State University, 525 Davey Laboratory, University Park, PA 16802, USA}

\author[0000-0002-0048-2586]{Andrew Monson}
\affiliation{Steward Observatory, The University of Arizona, 933 N. Cherry Avenue, Tucson, AZ 85721, USA}

\author[0000-0002-5463-9980]{Arvind Gupta}
\affiliation{U.S. National Science Foundation National Optical-Infrared Astronomy Research Laboratory, 950 N. Cherry Ave., Tucson, AZ 85719, USA}

\author[0000-0001-9662-3496]{William Cochran}
\affiliation{McDonald Observatory and Department of Astronomy, The University of Texas at Austin}
\affiliation{Center for Planetary Systems Habitability, The University of Texas at Austin}

\author[0000-0001-9596-7983]{Suvrath Mahadevan}
\affiliation{Department of Astronomy \& Astrophysics, The Pennsylvania State University, 525 Davey Laboratory, University Park, PA 16802, USA}

\author[0000-0003-4384-7220]{Chad Bender}
\affiliation{Steward Observatory, The University of Arizona, 933 N. Cherry Avenue, Tucson, AZ 85721, USA}

\author{Scott A. Diddams}
\affiliation{Electrical, Computer \& Energy Engineering, University of Colorado, 1111 Engineering Dr., Boulder, CO 80309, USA}
\affiliation{Department of Physics, University of Colorado, 2000 Colorado Avenue, Boulder, CO 80309, USA}

\author[0000-0003-1312-9391]{Samuel Halverson}
\affiliation{Jet Propulsion Laboratory, 4800 Oak Grove Drive, Pasadena, CA 91109, USA}

\author[0000-0002-9082-6337]{Andrea S.J. Lin}
\affiliation{Department of Astronomy, California Institute of Technology, 1200 E California Blvd, Pasadena, CA 91125, USA}

\author[0000-0002-0870-6388]{Maxwell Moe}
\affiliation{Department of Physics \& Astronomy, University of Wyoming, Laramie, WY 82070, USA}

\author[0000-0001-8720-5612]{Joe Ninan}
\affiliation{Department of Astronomy and Astrophysics, Tata Institute of Fundamental Research, Homi Bhabha Road, Colaba, Mumbai 400005,
India}

\author[0000-0003-0149-9678]{Paul Robertson}
\affiliation{Department of Physics \& Astronomy, The University of California, Irvine, Irvine, CA 92697, USA}


\author[0000-0001-8127-5775]{Arpita Roy}
\affiliation{Astrophysics \& Space Institute, Schmidt Sciences, New York, NY 10011, USA}

\author[0000-0002-4046-987X]{Christian Schwab}
\affiliation{School of Mathematical and Physical Sciences, Macquarie University, Balaclava Road, North Ryde, NSW 2109, Australia}

\author[0000-0001-7409-5688]{Gudmundur Stefansson}
\affiliation{Anton Pannekoek Institute for Astronomy, University of Amsterdam, Science Park 904, 1098 XH Amsterdam, The Netherlands}

\begin{abstract}

Brown dwarfs bridge the gap between stars and planets, providing valuable insight into both planetary and stellar formation mechanisms. Yet the census of transiting brown dwarf companions, in particular around M dwarf stars, remains incomplete. We report the discovery of two transiting brown dwarfs around low-mass hosts using a combination of space- and ground-based photometry along with near-infrared radial velocities. We characterize TOI-5389Ab ($68.0^{+2.2}_{-2.2} \ \mj$) and TOI-5610b ($40.4^{+1.0}_{-1.0} \ \mj$), two moderately massive brown dwarfs orbiting early M dwarf hosts  ($\teff = 3569 \pm 59 \ K$ and $3618 \pm 59 \ K$, respectively). For TOI-5389Ab, the best fitting parameters are period $P=10.40046 \pm 0.00002$ days, radius $R_{\rm BD}=0.824^{+0.033}_{-0.031}$~\rj, and low eccentricity $e=0.0962^{+0.0027}_{-0.0046}$. In particular, this constitutes one of the most extreme substellar-stellar companion-to-host mass ratios of $q=0.150$. For TOI-5610b, the best fitting parameters are period $P=7.95346 \pm 0.00002$ days, radius $R_{\rm BD}=0.887^{+0.031}_{-0.031}$ \rj, and moderate eccentricity $e=0.354^{+0.011}_{-0.012}$. Both targets are expected to have shallow but potentially observable secondary transits: $\lesssim 500$ ppm in Johnson K band for both. A statistical analysis of M-dwarf/BD systems reveals for the first time that those at short orbital periods ($P < 13$ days) exhibit a dearth of $13 \mj < M_{\rm BD} < 40 \mj$ companions ($q$ $<$ 0.1) compared to those at slightly wider separations.

\end{abstract}

\keywords{Brown dwarfs (185); Photometry (1234); Radial velocity (1332);
Spectroscopy (1558); Substellar companion stars (1648); Transit photometry (1709)
}

\section{Introduction}
\label{sec:intro}


Statistically, M dwarfs have a higher occurrence rate of small planets and lower occurrence rate of large planets compared to solar-type stars \citep{johnson2010, Bonfils2013-giantplanetsMdwarfs, mdwarf_planet_stats, Dressing/Charbonneau2015, hardegree-ulman, Hsu2020, Gan2023-MDplanets, Bryant2023-MDplanets}. 
As the most common stars in the Galaxy \citep{Henry2006-Mdwarfscommon, Reyle2021-Mdwarfscommon}, M dwarfs provide the greatest sample size to find rare systems in order to probe our understanding of planet formation.

Brown dwarfs (BDs) are defined as objects massive enough to burn deuterium but not massive enough that thermonuclear processes dominate their evolution
\citep[13 $\lesssim M\lesssim$ 80 \mj;][]{Burrows01, Lecavelier2022-IAU}. They are rare companions to main sequence (specifically FGK) stars, having $\lesssim 1\%$ occurrence \citep{BDdesert, Sahlman2011-BDdesert, santerne2016-BDdesert, Grieves2016-BDdesert}. The paucity of bound, short period BDs prompted the definition of the ``brown dwarf desert'' where these objects are exceedingly rare within $a\lesssim$~3~au of the stellar host, or equivalently $P \lesssim$ 5~yr \citep{BDdesert}.
 Additionally, the census of transiting BDs around M dwarfs is exceedingly small, with only 10 confirmed \citep{grieves_bdmd, canas_bd, henderson_bdmd}. These systems in particular are useful because they probe formation mechanisms and the transition from substellar to stellar objects, and they provide a more accurate measurement of the mass as opposed to $m \sin i$.
 
There is some debate on classifying planets and substellar objects by formation mechanism rather than mass \citep[i.e., core accretion vs gravitational instability;][]{santos2017-twopops, maldonado, kevin, kiefer2021}. Core accretion \citep{Mizuno1980-coreaccretion, Pollack1996-CoreAccretion, Chabrier2014-BDformation} involves planetesimals colliding and combining until they reach a critical mass ($\sim 10-30 \ M_\Earth$) beyond which they rapidly accrete gas in order to form giant planets. Gravitational instability \citep{Boss1997} involves objects forming instead by a secondary gravitational collapse of the protoplanetary disk or the cloud. 
Thus an alternative definition would stipulate that if bodies form via core accretion they are considered planets and if they form via gravitational instability (disk instability) they are considered BDs (unless they are massive enough to fuse hydrogen) \citep{Chabrier2014-BDformation}. Our targets specifically are so massive they much more likely formed via gravitational instability, since M dwarfs are observed to have lower mass protoplanetary disks compared to solar-type stars \citep{andrews2013-MassDiskRelation, Pascucci2016-MassDiskRelation, Boss2006}, making it harder to form large planets via core accretion \citep{kanodia2024GEMS}.
The definition of BDs based on their formation mechanism blurs the lower limit of 13 \mj. However, these more massive BDs still have unclear origins in if they formed via cloud or disk instabilities. 


The \textit{Searching for Giant Exoplanets around M-dwarf Stars (GEMS)} survey \citep{kanodia2024GEMS} utilizes space- and ground-based photometric and spectroscopic observations to target M-dwarfs within 200 pc to find GEMS. This survey is designed to characterize 40 GEMS in order to accurately compare GEMS to the existing census of hot-Jupiters around FGK stars. This is useful to constrain theories of planet formation. During the process of confirming and characterizing planets, many targets are observed to be brown dwarfs or binaries rather than planets, which is the case for the systems described in this work.

In this paper, we report the discovery and characterization of TOI-5389Ab and TOI-5610b, two BDs around M dwarfs originally flagged as objects of interest by the Transiting Exoplanet Survey Satellite \citep[TESS;][]{Ricker15} science team. Section \ref{sec: pbservations} presents new ground-based photometric and spectroscopic data on these targets. Section \ref{sec: analysis} describes the combination and modeling of these datasets to extract planetary, stellar, and system parameters. Section \ref{sec: discussion} provides a discussion of the new data in the context of other transiting BD systems. Finally, Section \ref{sec: Conclusion} summarizes and concludes our findings.

\section{Observations} \label{sec: pbservations}
\subsection{TESS Photometry}

TOI-5389A (TIC 39143128, Gaia EDR3 765082727962822016) was identified via the TESS Science Processing Operations Center \citep[SPOC;][]{Jenkins2016-TESS_SPOC} and was observed in 1800~s cadence in Sector 22 from 2020 February 18 to 2020 March 18 and Sector 48 in 600~s cadence from 2022 January 28 to 2022 February 26. TOI-5610 (TIC 252481136, Gaia EDR3 834714838005307648) was identified via the TESS Faint Star Search \citep{Kunimoto2022-TESSfaintstar} and was observed in Sector 21 in 1800~s cadence from 2020 January 21 to 2020 February 18 and Sector 48 in 600~s cadence from 2022 January 28 to 2022 February 26.


To produce a dilution-free lightcurve, we used the {\tt TESS Gaia Light-Curve} \citep[TGLC;][]{Han2023-TGLC} python package, which de-blends the lightcurve using Gaia DR3 \citep{gaia2016, gaiadr3-2023} photometry and positions. Within the 3x3 TESS pixel aperture used by TGLC, a Gaia DR3 search using a $\sim 89\arcsec$ radius revealed that TOI-5389A has five potential nearby contaminant stars which could cause dilution in the lightcurve depending on target placement within the pixel. The collective targets have a Gaia $\Delta g$ mag range of $\sim$ 1--5 mag. Similarly, TOI-5610 also has five potential contamination sources with a Gaia $\Delta g$ mag range of $\sim$ 1--5 mag. Only TESS Sector 48 for TOI-5610 appeared to have appreciable dilution from another source, the others either had more favorable pixel placement or negligible dilution. TGLC corrects for most of this, and we also constrain any residual dilution using other un-diluted ground-based lightcurves (Section \ref{sec: RBO}).

There are clear transit signatures at dominant periodicities of $\sim$10.40 and $\sim$7.95 days and approximate depths of $\sim$ 4--5\% and $\sim$ 2--3\% for TOI-5389A and TOI-5610, respectively. A Box Least Squares analysis of both targets shows no other significant periodicities.

\begin{figure*}
    \centering
    \includegraphics[width=\textwidth]{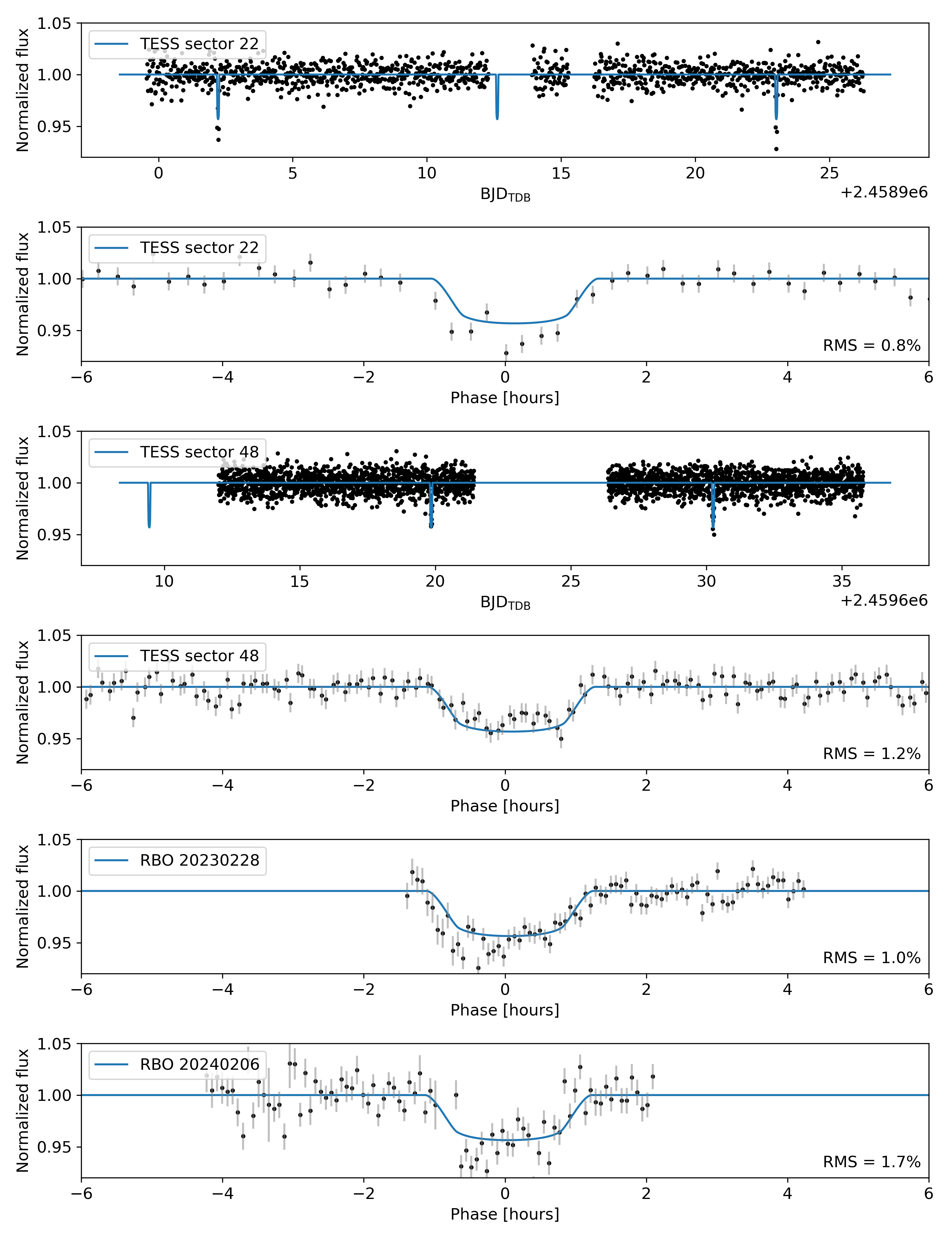}
    \caption{Observed transits for TOI-5389Ab. From first to last: TESS Sector 22 (2020 March) and its phase folded counterpart, TESS Sector 48 (2022 February) and its phase folded counterpart, RBO I-band 2023 February 28, and RBO I-band 2024 February 06.}
    \label{fig: all transits}
\end{figure*}

\begin{figure*}
    \centering
    \includegraphics[width=\textwidth]{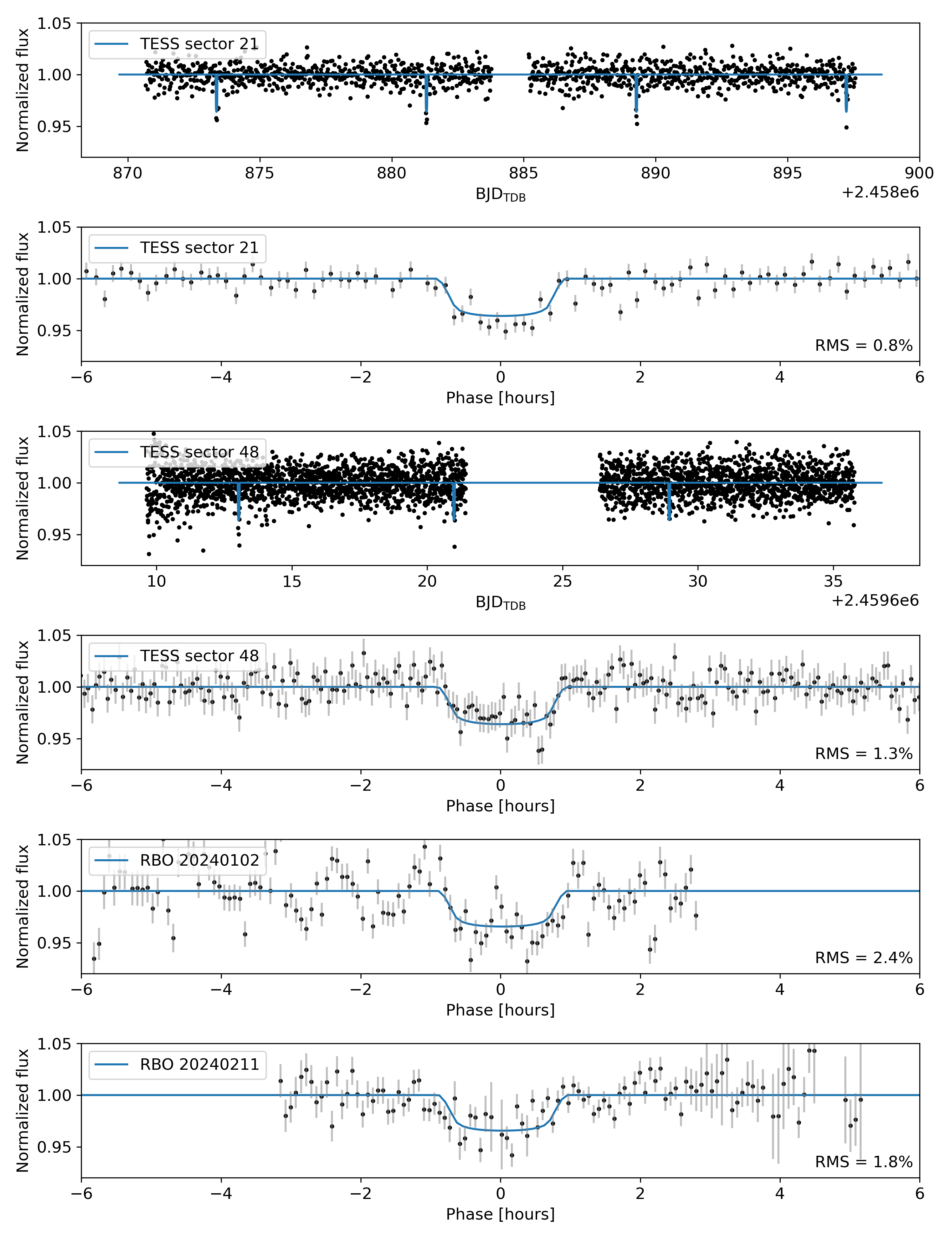}
    \caption{Observed transits for TOI-5610b. From first to last: TESS Sector 21 (2020 February) and its phase folded counterpart, TESS Sector 48 (2022 February) and its phase folded counterpart, RBO I-band 2024 January 02, and RBO I-band 2024 February 11.}
    \label{fig: all transits 5610}
\end{figure*}

Figure~\ref{fig: all transits} shows the two full TESS Sector lightcurves and their phase-folded counterparts for TOI-5389A ({\it black dots}) and Figure \ref{fig: all transits 5610} shows the same for TOI-5610. The blue curves shows the best-fit models, described in Section \ref{sec: analysis}. 

\subsection{Red Buttes Observatory Photometry}
\label{sec: RBO}

The Red Buttes Observatory \citep[RBO;][]{Kasper16} is a 0.6~m telescope owned by the University of Wyoming located 10 km south of Laramie, Wyoming. It is an f/8.43 Ritchey-Cretien by DFM Engineering, Inc. equipped with an Apogee Alta F16 camera. The Kodak KAF 16801 4096$\times$4096 chip with 9 micron pixels binned 2$\times$2 produces a plate scale of 0.73 arcsec pix$^{-1}$ and a field of view of 24.9 arcmin. All observations were conducted using the Bessell I filter and a 240~s exposure time while slightly defocused. The CCD has a gain of 1.39 $e^-$ ADU$^{-1}$ and was operated at $-15$ C which results in a significant dark current ($\sim 0.13 \ e^- \ s^{-1} \ pix^{-1}$). Exposures were calibrated using flat-field, bias, and dark corrections. The readout time was $\sim$2.4~s and a typical full width at half maximum is $\sim 3 \arcsec$. The typical photometric uncertainty is 0.9--1.2\%. We successfully observed TOI-5389Ab on the local nights of 2023 February 28 and 2024 February 6 as well as TOI-5610b on the local nights of 2024 January 2 and 2024 February 11. Observations lasted 5--8 hours each and captured two full transits for each target. Observation times had to be converted\footnote{\url{https://astroutils.astronomy.osu.edu/time/utc2bjd.html}} from JD$_{\rm UTC}$ into \bjdtdb \ as described in \cite{eastman2010-BJDTDB}. The lightcurves were derived from the raw data via differential aperture photometry in AstroImageJ \citep{Collins2017-AIJ}. The final two panels of Figures \ref{fig: all transits} \& \ref{fig: all transits 5610} display the normalized RBO lightcurves for each epoch. Tables \ref{tab: RBO photometry} and \ref{tab: RBO phot 5610} present the Julian dates, normalized flux, and uncertainties. The full tables are available as machine-readable files. 

\begin{table}[h]
    \centering
    \caption{TOI-5389 RBO Photometric Data}
    \begin{tabular}{c c c}
    \hline
    \hline
    \bjdtdb & Normalized Flux & Flux err \cr
    \hline
    2460004.605 & 0.995 & 0.013 \cr
    2460004.608 & 1.018 & 0.013 \cr
    2460004.611 & 1.011 & 0.013 \cr
    ... & ... & ... \cr
    \hline
    \end{tabular}
    \label{tab: RBO photometry}
\end{table}

\begin{table}[h]
    \centering
    \caption{TOI-5610 RBO Photometric Data}
    \begin{tabular}{c c c}
    \hline
    \hline
    \bjdtdb & Normalized Flux & Flux err \cr
    \hline
    2460312.694 & 0.934 & 0.016 \cr
    2460312.697 & 0.949 & 0.016 \cr
    2460312.700 & 0.999 & 0.016 \cr
    ... & ... & ... \cr
    \hline
    \end{tabular}
    \label{tab: RBO phot 5610}
\end{table}

\subsection{NESSI High Resolution Speckle Imaging}

\begin{figure}[h!]
    \centering
    \includegraphics[scale=0.27]{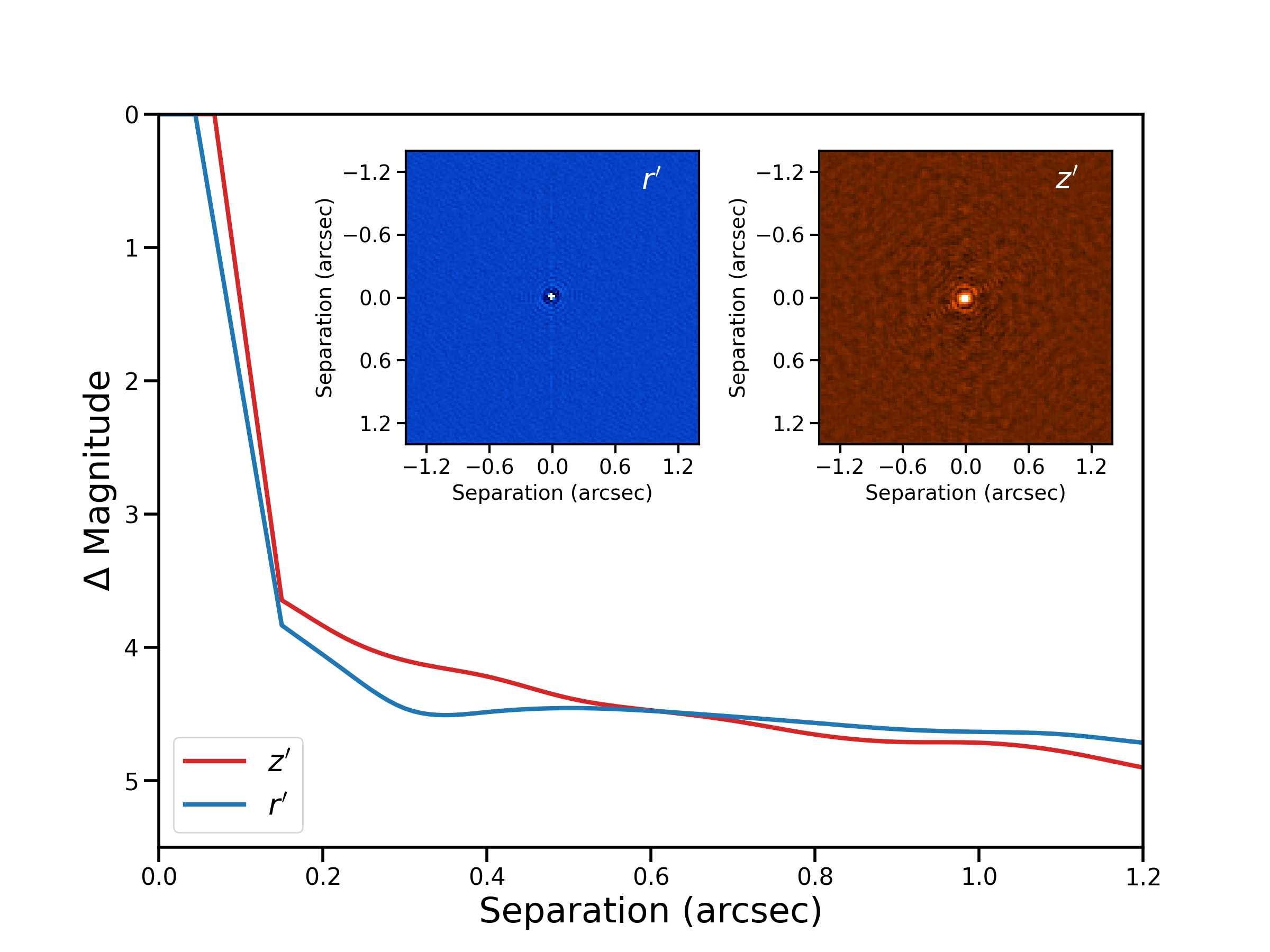}
    \caption{NESSI contrast curves for TOI-5389A. NESSI rules out nearby sources down to  $\Delta r' = 4.05$ mag and $\Delta z' = 3.84$ mag within 0\farcs{2}.}
    \label{fig: 5389NESSI}
\end{figure}

\begin{figure}[h!]
    \centering
    \includegraphics[scale=0.27]{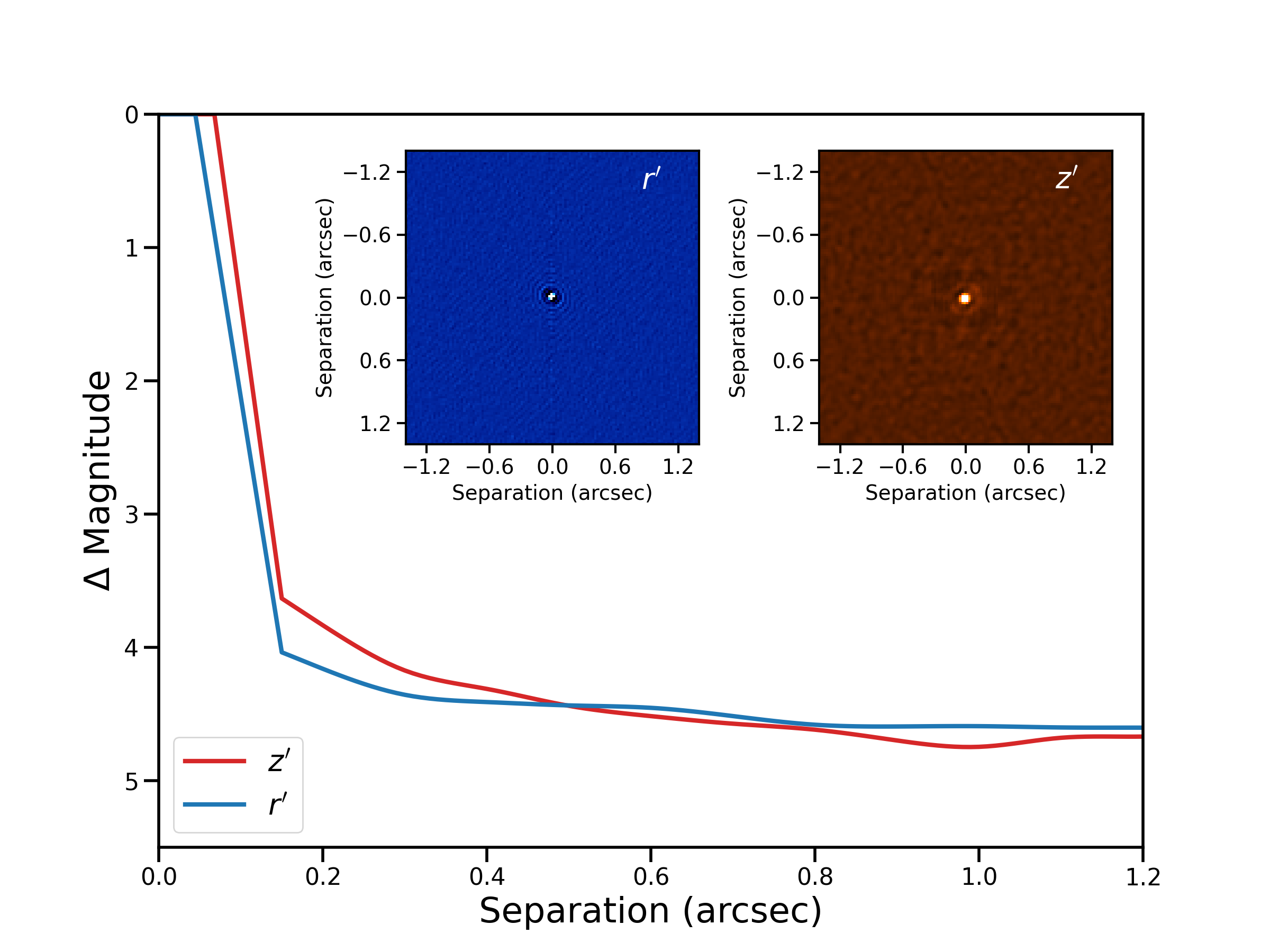}
    \caption{NESSI contrast curves for TOI-5610. NESSI rules out nearby sources down to  $\Delta r' = 4.16$ mag and $\Delta z' = 3.83$ mag within 0\farcs{2}.}
    \label{fig: 5610NESSI}
\end{figure}

TOI-5389A and TOI-5610 were observed with the NN-Explore Exoplanet Stellar Speckle Imager \citep[NESSI;][]{Scott2018-NESSI} on the WIYN 3.5~m telescope\footnote{The WIYN Observatory is a joint facility of the NSF's National Optical-Infrared Astronomy Research Laboratory, Indiana University, the University of Wisconsin-Madison, Pennsylvania State University, Purdue University and Princeton University.} on 2023 January 28. For each target, a set of diffraction-limited frames was taken simultaneously with the red and blue NESSI cameras in the Sloan Digital Sky Survey \citep[SDSS;][]{Kollmeier2019-SDSS} {\it z'} and {\it r'} filters. The data were processed following the methods described in \cite{Howell2011-NESSI}. Figures \ref{fig: 5389NESSI} and \ref{fig: 5610NESSI} show the reconstructed speckle images and contrast limits. For TOI-5610, these data allow us to rule out the presence of nearby sources down to magnitude limits of $\Delta r' = 4.16$ mag and $\Delta z' = 3.83$ mag at a separation of 0\farcs{2} and limits of $\Delta r' = 4.60$ mag and $\Delta z' = 4.67$ mag at a separation of 1\farcs{2}. For TOI-5389A, the equivalent magnitude limits are $\Delta r' = 4.05$ mag and $\Delta z' = 3.84$ mag at a separation of 0\farcs{2} and $\Delta r' = 4.72$ mag and $\Delta z' = 4.90$ mag at a separation of 1\farcs{2}.
The Gaia RUWE values of 1.147 and 0.999 for TOI-5389A and TOI-5610, respectively, are consistent with single star astrometric solutions \citep[RUWE $>$ 1.4 is suggestive of binarity;][]{Belokurov2020-ruwe, Penoyre2020-ruwe}. The GAIA Astrometric Excess Noise is also consistent with this.

\subsection{The Habitable-zone Planet Finder Spectra}

The Habitable-zone Planet Finder \citep[HPF;][]{mahadevan2012hpf, mahadevan2014hpf} is a high-resolution, near infrared, fiber-fed \citep{kanodia2018hpffibers}, stabilized \citep[$\sim$ 1 mK;][]{stefansson2016stabilized} spectrograph with resolution $R \sim 55,000$ and wavelength coverage 8080--12780 \AA. It is located on the 10~m Hobby-Eberly Telescope \citep{ramseyHET, Hill2021-HET} at the McDonald Observatory in Texas. We obtained two 945~s exposures on each of ten nights for TOI-5389A and twelve nights for TOI-5610 between 2022 November and 2023 December.

The raw data were processed using the {\tt HxRGproc} algorithms \citep{Ninan2018-hgxcproc} and wavelength calibrated by the method described in \cite{hpfspecmatch}. Wavelength solutions were determined from routine laser frequency comb (LFC) calibration frames that provide a drift correction to a precision of $<$ 30 cm s$^{-1}$ \citep{hpfspecmatch}. These were not simultaneous with observations to avoid scattered light in the spectra. The extracted 1D spectra have a median signal-to-noise ratios (S/N) per pixel of 18 (TOI-5389A) and 15 (TOI-5610) in spectral order 18 at $1 \ \mu m$.


The 28 spectral orders were continuum normalized and combined into a single spectrum which was then Doppler corrected to the barycentric frame of reference using {\tt barycorrpy}, a python transposition of the \cite{Wright-Eastman2014-barycorr} algorithms \citep{Kanodia-wright2018-barycorrpy}. Wavelengths in regions of poor atmospheric transmission were masked. Only the wavelength intervals 8600--8900 \AA\ and 9900--10600 \AA\ were retained for radial velocity analysis. We employed a custom python code to compute the broadening function \citep[analogous to a cross correlation;][]{Rucinski1999} using a high-resolution PHOENIX \citep{Husser2013} model atmosphere of the appropriate effective temperature and gravity for each target. Radial velocities were measured as the center of a Gaussian function fitted to the peak in the broadening function at each of the observed epochs. Uncertainties on the Gaussian center yield a typical precision of 20--30 m s$^{-1}$ via a least squares fit to the cross-correlation function. A test of this analysis technique on HPF spectra of Barnard's star reproduces the accepted literature velocities within 0.3 \kms, providing an overall velocity accuracy measurement \citep[GJ699, an M dwarf having a well-established radial velocity of -110.1 \kms;][]{Fouque2018}.  While, in principle, the rotational velocities of the target stars are measurable from the width of the broadening function after deconvolution with the instrumental broadening, the $vsini$ for these spectra are all less than 2 \kms, below the instrumental limit set by HPF's spectral resolution of $R \approx 55,000$. These slower rotational velocities are consistent with the slow rotational velocities expected of older early M-dwarfs \citep{Giacobbe2020-MdwarfRotation}.

\begin{figure}
    \centering
    \includegraphics[width=\linewidth]{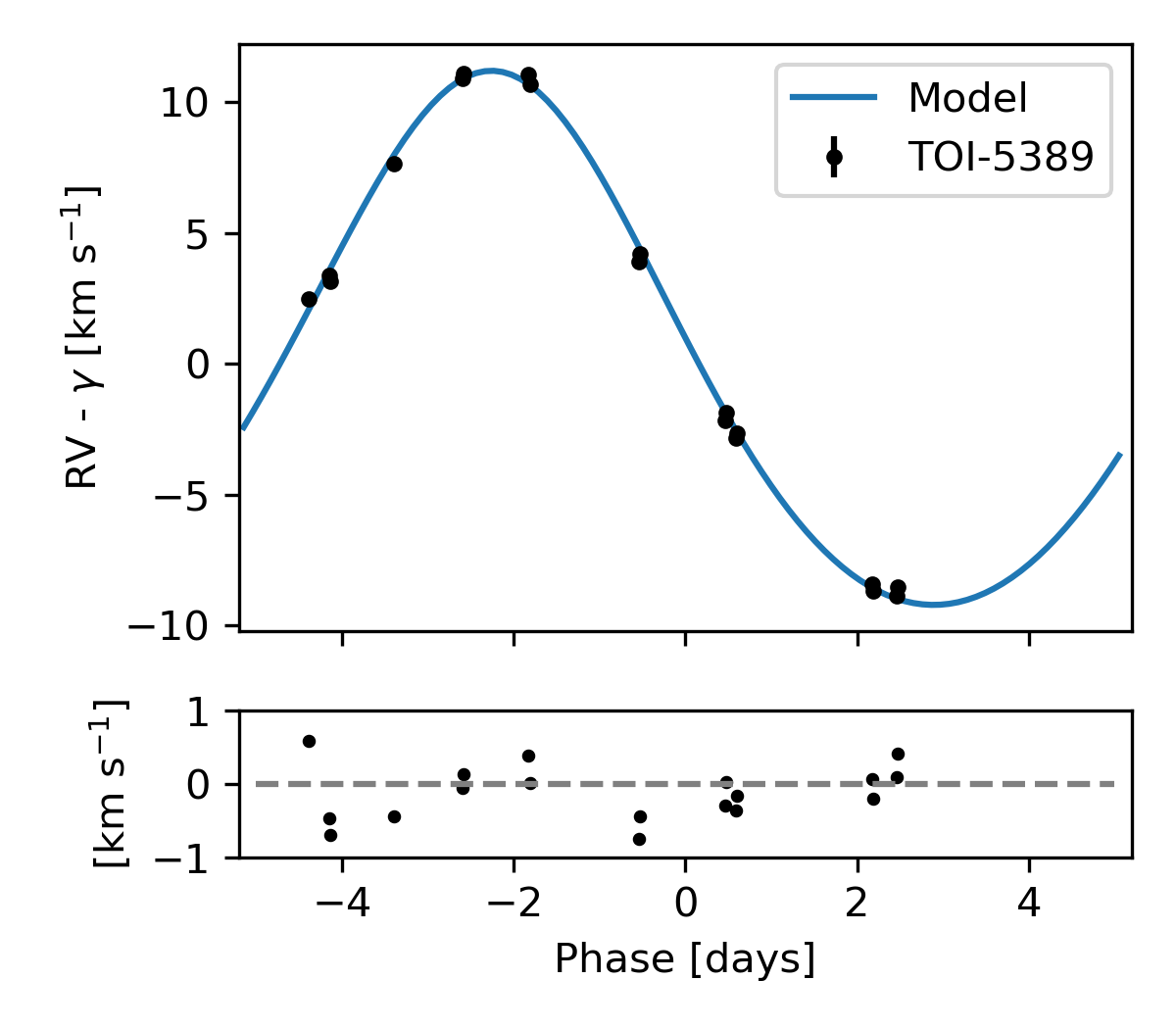}
    \caption{Phased barycentric radial velocity curve and best-fit model for TOI-5389A. Residuals are plotted below. Systemic velocity $\gamma$ is subtracted. Errorbars are plotted but are too small to see.}
    \label{fig: rv}
\end{figure}

\begin{figure}
    \centering
    \includegraphics[width=\linewidth]{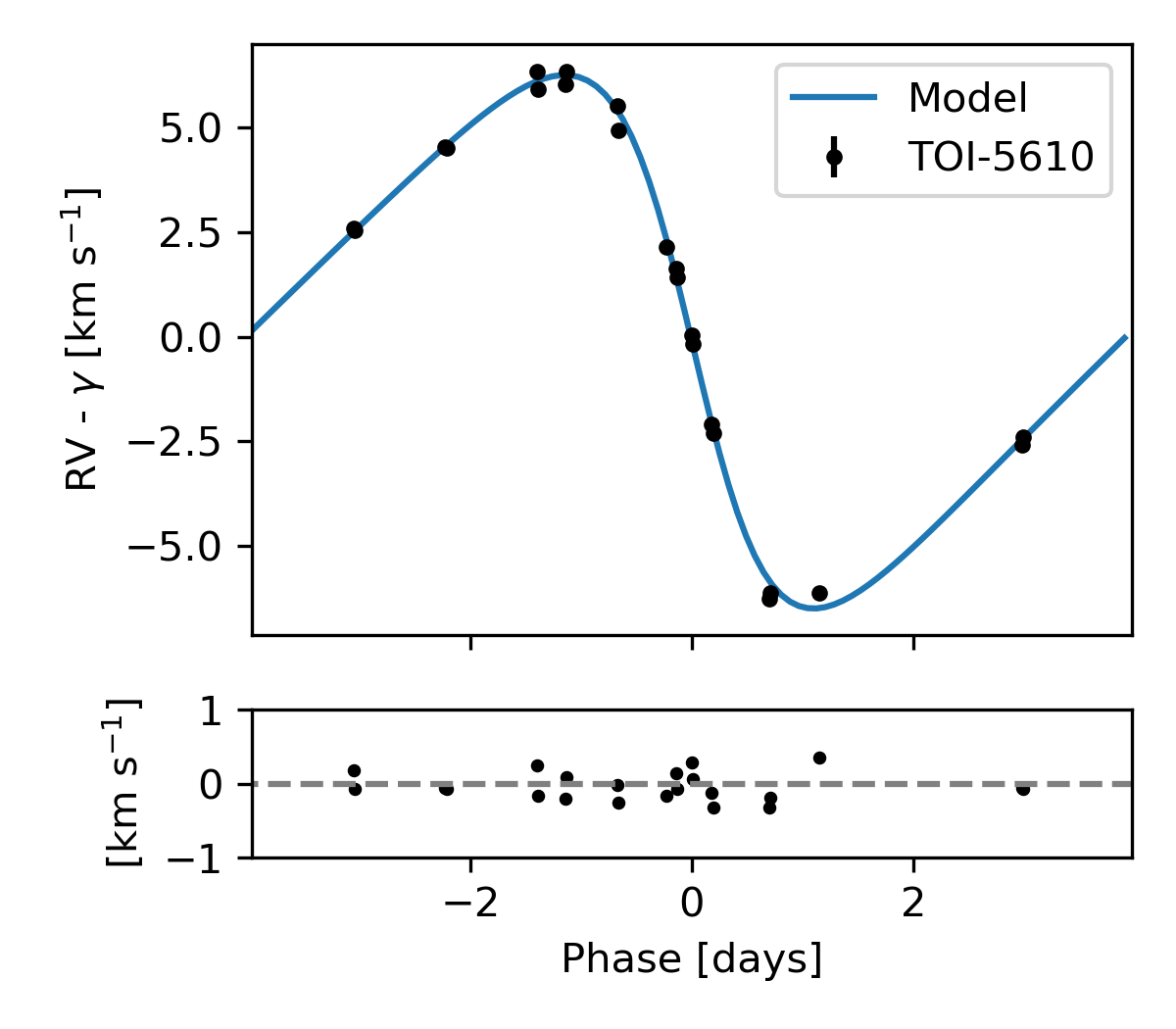}
    \caption{Phased barycentric radial velocity curve and best-fit model for TOI-5610. Residuals are plotted below. Systemic velocity $\gamma$ is subtracted. Errorbars are plotted but are too small to see.}
    \label{fig: rv5610}
\end{figure}

Tables \ref{tab: rv table 5389} and \ref{tab: rv table 5610} list the HPF RVs, while Figures \ref{fig: rv} \& \ref{fig: rv5610} show the phased RV time series and residuals. For both targets, the unphased velocity curve is consistent with uncertainties and the residuals show no long term patterns.

\begin{table}[h]
    \centering
    \caption{Barycentric Radial Velocity Measurements for TOI-5389A}
    \begin{tabular}{c c c}
    \hline
    \hline
    \bjdtdb & RV (m s$^{-1}$) & $\sigma_{RV}$ (m s$^{-1}$) \cr
    \hline
    2459921.927 & 9123 & 23 \cr
    2459921.939 & 9443 & 23 \cr
    2460306.864 & 8455 & 16 \cr
    2460306.876 & 8654 & 15 \cr
    2459954.830 & 2884 & 16 \cr
    2459954.842 & 2607 & 19 \cr
    2459923.921 & 2424 & 24 \cr
    2459923.933 & 2752 & 30 \cr
    2460301.884 & 13774 & 81 \cr
    2460010.912 & 14673 & 17 \cr
    2460010.924 & 14447 & 28 \cr
    2460302.877 & 18928 & 34 \cr
    2460095.676 & 22214 & 20 \cr
    2460095.688 & 22399 & 17 \cr
    2459950.835 & 22354 & 43 \cr
    2459950.847 & 21974 & 27 \cr
    2460284.933 & 15185 & 24 \cr
    2460284.945 & 15503 & 20 \cr
    \hline
    \end{tabular}
    \label{tab: rv table 5389}
\end{table}

\begin{table}[h]
    \centering
    \caption{Barycentric Radial Velocity Measurements for TOI-5610}
    \begin{tabular}{c c c}
    \hline
    \hline
    \bjdtdb & RV (m s$^{-1}$) & $\sigma_{RV}$ (m s$^{-1}$) \cr
    \hline
    2459931.869 & 37125 & 14 \cr
    2459931.881 & 37258 & 18 \cr
    2460011.860 & 37254 & 23 \cr
    2460029.596 & 40797 & 6 \cr
    2460029.608 & 41000 & 5 \cr
    2460007.649 & 45996 & 9 \cr
    2460007.661 & 45944 & 9 \cr
    2459952.803 & 47922 & 7 \cr
    2459952.815 & 47900 & 8 \cr
    2459897.954 & 49727 & 13 \cr
    2459897.966 & 49320 & 16 \cr
    2460303.842 & 49440 & 22 \cr
    2460303.854 & 49730 & 19 \cr
    2460065.705 & 48915 & 12 \cr
    2460065.717 & 48339 & 12 \cr
    2460272.939 & 45533 & 13 \cr
    2460304.841 & 45033 & 9 \cr
    2460304.853 & 44817 & 10 \cr
    2459955.029 & 43441 & 12 \cr
    2459955.041 & 43212 & 11 \cr
    2460018.840 & 41292 & 6 \cr
    2460018.852 & 41093 & 7 \cr
    \hline
    \end{tabular}
    \label{tab: rv table 5610}
\end{table}

\section{Analysis}
\label{sec: analysis}

\subsection{Stellar Parameters}
\label{sec: stellar parameters}

Table \ref{tab: literature stellar parameters} lists broadband photometric measurements from the optical through mid-infrared from the Panoramic Survey Telescope and Rapid Response System \citep[Pan-STARRS;][]{chambers2016-pstarrs}, the 2-Micron All Sky Survey \citep[2MASS;][]{Shrutskie2006-2MASS}, and the Wide-field Infrared Survey Explorer \citep[WISE;][]{Wright2010-WISE}. Table \ref{tab: literature stellar parameters} also contains stellar parameters \teff, $\log{g}$, \feh, mass ($M_\star$), radius ($R_\star$), parallax (Plx), RA proper motion (RA PM), DEC proper motion (DEC PM), density ($\rho_\star$), luminosity (L), and spectral type (Sp-Type). With spectra from HPF, we used the spectral matching technique ({\tt HPF-SpecMatch}) described in \cite{hpfspecmatch} to independently obtain stellar parameters \teff, \logg, and \feh. We selected the night with the highest S/N spectrum in the 5th spectral order ($\lambda$$\sim$8670 \AA--8750 \AA) for the least amount of telluric contamination. {\tt HPF-SpecMatch} then compares this spectrum to a library of high S/N ($>$ 100) spectra in order to derive new stellar parameters. The library of 166 stars spans $2700 $ K$< \teff < 6000$ K, $4.3 <$ \logg $\ < 5.3$, and $-0.5 < \feh < 0.5$. The TOI-5610 metallicity estimate is close to this lower limit.

It should be noted that the metallicity estimates from {\tt HPF-SpecMatch} entail some caveats. Since M dwarfs are lower temperatures and have atmospheres characterized by molecular features, their metallicities are notoriously difficult to determine \citep{passeger2022-MDwarfmetallicity}. In addition, {\tt HPF-SpecMatch} provides only loose constraints on metallicity because the $\chi^2$ suffers from some multi-modality. The \teff \ and \logg \ estimates, however, are considered accurate. The errors for these quantities come from cross validation comparing values recovered from {\tt HPF-SpecMatch} with library values.

Table \ref{tab: literature stellar parameters} includes these best fitting stellar parameters of \teff = 3569 $\pm$ 59 K, \logg = 4.79 $\pm$ 0.04, and \feh = -0.15 $\pm$ 0.16 for TOI-5389A, and parameters of \teff = 3618 $\pm$ 59 K, \logg = 4.78 $\pm$ 0.04, and \feh = -0.42 $\pm$ 0.16 for TOI-5610. The temperatures and gravities of both targets correspond to M1V--M2V stars as defined by \cite{Boyajian2012spectraltypes}. 

\begin{table*}[h]
    \centering
    \caption{Stellar parameters.}
    \begin{tabular}{c c c c}
    \hline
    \hline
    Parameter & TOI-5389A & TOI-5610 & Source \cr
    \hline
    RA & 11:15:12.4 & 10:24:05.48 & Gaia \cr
    DEC & +39:21:32.11 & +48:14:53.99 & Gaia \cr
    RA PM (mas yr$^{-1}$) & $-33.0227 \pm 0.0438$ & $-27.3901 \pm 0.0328$ & Gaia \cr
    DEC PM (mas yr$^{-1}$) & $-46.7001 \pm 0.0404$ & $-55.2985 \pm 0.0316$ & Gaia \cr
    \hline
    g & $17.526 \pm 0.015$ & $17.289 \pm 0.014$ & Pan-STARRS \cr
    r & $16.310 \pm 0.008$ & $16.153 \pm 0.010$ & Pan-STARRS \cr
    i & $15.26 \pm 0.007$ & $15.290 \pm 0.013$ & Pan-STARRS \cr
    z & $14.796 \pm 0.016$ & $14.896 \pm 0.010$ & Pan-STARRS \cr
    J & $13.402 \pm 0.024$ & $13.598 \pm 0.023$ & 2MASS \cr
    H & $12.81 \pm 0.018$ & $12.964 \pm 0.024$ & 2MASS \cr
    K & $12.555 \pm 0.023$ & $12.781 \pm 0.026$ & 2MASS \cr
    W1 & $12.461 \pm 0.024$ & $12.674 \pm 0.023$ & WISE \cr
    W2 & $12.323 \pm 0.024$ & $12.542 \pm 0.024$ & WISE \cr
    W3 & $12.016 \pm 0.327$ & $12.01 \pm 0.272$ & WISE \cr
    \hline
    \teff (K) & $ 3569 \pm 59$ & $3618 \pm 59$ & This work \cr
    \logg &  $4.79 \pm 0.04$ & $4.78 \pm 0.04$ & This work \cr
    $M_\star$ (\msun) & $0.43 \pm 0.02$ & $0.53 \pm 0.02$ & This work \cr
    $R_\star$ (\rsun) & $0.42 \pm 0.02$ & $0.52 \pm 0.02$ & This work \cr
    \feh\footnote{These numbers entail some caveats. See Section \ref{sec: stellar parameters}.} & $-0.15 \pm 0.16$ & $-0.42 \pm 0.16$ & This work \cr
    Parallax (mas) & $5.1298 \pm 0.057$ & $3.6509 \pm 0.037$ & Gaia \cr
    $\rho_\star$ (g cm$^{-3}$) & $8.19 \pm 1.23$ & $5.32 \pm 0.65$ & This work \cr 
    L (L$_{\odot}$) & $0.026 \pm 0.004$ & $0.042 \pm 0.005$ & This work \cr
    Sp-Type & M1V--M2V & M1V--M2V & This work \cr
    \hline
    \end{tabular}
    \label{tab: literature stellar parameters}
\end{table*}



We took these spectroscopic parameters and used them as priors in a spectral energy distribution (SED) fit using {\tt EXOFASTv2} \citep{Eastman2019-exofastv2} to obtain stellar mass and radius. {\tt EXOFASTv2} utilizes MIST \citep{Dotter2016-MIST, Choi2016-MIST} stellar evolutionary models and tracks in conjunction with NextGen \citep{allard2012-btsettl} stellar atmosphere models. In addition to stellar \teff, \logg, \feh, and the Gaia parallax, we used Pan-STARRS $griz$, 2MASS $JHK$, and WISE 1 and 2 photometric measurements as listed in Table \ref{tab: literature stellar parameters} as constraints. Table \ref{tab: literature stellar parameters} lists best-fit results of $M_\star = 0.43 \pm 0.02 \msun$ and $R_\star = 0.42 \pm 0.01 \rsun$ for TOI-5389A. It also lists best fit results of $M_\star = 0.53 \pm 0.02 \msun$ and $R_\star = 0.52 \pm 0.01 \rsun$ for TOI-5610. As an alternative method of radius estimation we used a Python program\footnote{doi:10.5281/zenodo.11401588} based on the \cite{Kiman2024-radius_est} method which uses surface brightness–color relations for the three Gaia magnitudes with Gaia DR3 parallaxes and found well-agreeing results of 0.422 $\pm$ 0.005 \rsun \ and 0.513 $\pm$ 0.006 \rsun \ for TOI-5389A and TOI-5610, respectively. We chose to use the EXOFASTv2 values since they are based on actual measurements of the star and the availability of the posteriors. 

\subsection{TOI-5389B: White Dwarf Companion}
\label{sec: WD}

\cite{elbadry2021-bindaries} lists TOI-5389A as a wide binary with the white dwarf star Gaia DR3 765082934121253376 (listed as spectral type DC, henceforth TOI-5389B). TOI-5389B has a G mag of 20.19, and the projected separation is 12.6 \arcsec,  corresponding to 2447 AU at a distance of 196 pc (TOI-5389A). Their respective Gaia DR3 parallaxes are 5.1298 $\pm$ 0.057 $mas$ and 7.6227 $\pm$ 1.2444 $mas$, which are in agreement near the $\sim 2 \sigma$ level. The catalog also reports that the likelihood of this being a chance alignment of proper motions is extremely small ($R \sim 10^{-9}$), implying that these objects constitute a bound system. The catalog does not contain TOI-5610, and so we assume no binarity.

Using DA and DB white dwarf cooling curves\footnote{\url{http://www.astro.umontreal.ca/~bergeron/CoolingModels}} based on \cite{holberg2006-WD, kowalski2006-WD, tremblay2011-WD, Blouin2018-WD, bedard2020-WD}, we fit the SED of TOI-5389B using a $\chi^2$ minimization method. We used the Gaia $G$, $BP$, $RP$, and SDSS $u$, $g$, $r$, $i$, $z$ magnitudes with the TOI-5389A parallax to find the best-fitting intrinsic white dwarf SED. The best fit model to TOI-5389B yields a spectral type DB with $M = 0.573 \msun$, $\teff=6500$ K, $logg = 7.0$, and a post-MS age of $\sim2$ Gyr. Therefore, TOI-5389A must be at least 2 Gyr old.

\subsection{Gyrochronology}
\label{sec: gyro}

Adopting an upper limit on $v \sin i$ of 2 \kms, corresponding to the minimum detectable rotational velocity of HPF, and using the radii of the stars, we estimate lower limits on the rotation periods of 11 and 13 days for TOI-5389A and TOI-5610, respectively. A Lomb-Scargle analysis of the TESS, Zwicky Transient Facility \citep[ZTF DR22;][]{masci2019-ZTF, bellm2019-ZTF}, and All Sky Automated Survey for SuperNovae \citep[ASAS-SN Sky Patrol v2.0;][]{Shappee2014-ASASSN, hart2023-ASASSN} data produced no identifiable stellar rotational period, so we proceed with our limiting estimates. \cite{Engle2018-gyrochronology} presents two relations for age as a function of rotational period of M dwarfs: one for M0V--M1V and another for M2.5V--M6V stars. Our targets fall between these two spectral ranges. For both stars and both relations, this suggests ages of $\gtrsim 1$ Gyr.
The lack of rotational broadening and photometric rotational modulation are indicative of an inactive star.

Separately, {\tt EXOFASTv2} yields SED-based ages of $10.0^{+2.6}_{-4.0}$ and $10.0^{+2.6}_{-3.9}$ Gyr for TOI-5389A and TOI-5610, respectively.

\subsection{Galactic Population Analysis}

\begin{table}[h!]
    \centering
    \begin{tabular}{c|c|c}
        \hline
        \hline
        Parameter & TOI-5389A & TOI-5610 \cr
        \hline
        U & -9.02 & 2.10 \cr
        U (LSR) & $2.08 \pm 0.17$ & $13.20 \pm 0.24$ \cr
        V & -51.31 & -82.81 \cr
        V (LSR) & $-39.07 \pm 0.60$ & $-70.57 \pm 0.78$ \cr
        W & -14.41 & -37.96 \cr
        W (LSR) & $-7.16 \pm 0.10$ & $-30.71 \pm 0.09$ \cr
        \hline
        $\mathcal{P}$ (thin) & $0.98 \pm 0.02$ & $0.69 \pm 0.05$ \cr
        $\mathcal{P}$ (thick) & $0.02 \pm 0.00003$ & $0.30 \pm 0.002$ \cr
        $\mathcal{P}$ (halo) & $0.00005 \pm 0.0000005$ & $0.002 \pm 0.00002$ \cr
        \hline
    \end{tabular}
    \caption{Galactic velocities and probabilities of being in the thin disk, thick disk, or halo.}
    \label{tab: UVW}
\end{table}

Table \ref{tab: UVW} shows galactic velocities in barycentric and Local Standard of Rest \citep[LSR;][]{Schonrich2010-LSR} reference frames as well as the probabilities of being associated with the thin disk, thick disk, or halo populations. Based on these results, TOI-5610 could belong to either the galactic thin or thick disk. This agrees well with our metallicity estimates, and the velocities allow for the possibility of being an old star. \citep{Holmberg2019, Sharma2014, Hwang2020}. TOI-5389A has kinematics consistent with thin disk stars, which agrees with our metallicity estimates but allows a very wide range of ages.

\subsection{Physical and Orbital Parameters}

We also used {\tt EXOFASTv2} to derive the companions' physical properties and orbital parameters by jointly modelling the four lightcurves and the radial velocity curve of each target. It should be noted that this fit also included an SED fit, which reproduced the initial values. We adopted normal stellar priors from Table \ref{tab: literature stellar parameters} along with loose uniform priors on extinction, eccentricity, equivalent evolutionary phase (EEP), systemic velocity, and system inclination. We required the default criteria for convergence given by {\tt EXOFASTv2} which is a Gelman-Rubin statistic $<1.01$ and $tz$ (number of independent chains) $>1000$. We examined the Probability Distribution Functions (PDFs) to ensure they were approximately Gaussian and that all chains converge on the same solution. Table \ref{tab: final parameters} shows the best fitting final parameters with uncertainties for the companions and their orbits, including period (P), time of primary eclipse ($t_0$), impact parameter (b), secondary impact parameter (b$_s$), eccentricity ($e$), systemic velocity ($\gamma$), BD radius ($R_{\rm BD}$), BD mass ($M_{\rm BD}$), BD surface gravity ($\log{g_{\rm BD}}$), argument of periastron of the star ($\omega$), semi-major axis ($a$), inclination ($i$), time of secondary eclipse ($t_s$), Sonora Bobcat \citep{marley_bdatmosphere} model-based estimated BD temperature ($T_{SB}$, discussed in section \ref{sec: bobcat}), transit depth in Johnson I band, occultation depth in Johnson I band, and occultation depth in Johnson K band.
For TOI-5389Ab we find a mass of 68.0$^{+2.2}_{-2.2}$ \mj---near the upper mass limit for BDs---, a radius of 0.824$^{+0.033}_{-0.031}$ \rj, and an eccentricity of 0.096$^{+0.003}_{-0.005}$.
For TOI-5610b we find a mass of 40.4$^{+1.0}_{-1.0}$ \mj, a radius of 0.887$^{+0.031}_{-0.031}$ \rj, and a moderate eccentricity of 0.354$^{+0.011}_{-0.012}$.

\begin{table}[h]
    \centering
    \caption{Physical and Orbital Parameters}
    \begin{tabular}{c c c}
    \hline
    \hline
    Parameter & TOI-5389Ab & TOI-5610b \cr
    \hline
    P (days) & $10.40046 \pm 0.00002$ & $7.95346 \pm 0.00002$ \cr
    $t_0$ (\bjdtdb) & 2459609.445$^{+0.00096}_{-0.00091}$ & 2459628.939$^{+0.0013}_{-0.0012}$ \cr
    b ($R_\star$) & 0.386$^{+0.09}_{-0.13}$ & 0.124$^{+0.12}_{-0.09}$ \cr
    b$_s$ ($R_\star$) & 0.396$^{+0.08}_{-0.13}$ & 0.26$^{+0.24}_{-0.18}$ \cr
    $e$ & 0.096$^{+0.003}_{-0.005}$ & 0.354$^{+0.011}_{-0.012}$ \cr
    $\gamma$ (m/s) & -11300$^{+110}_{-100}$ & -43387$^{+65}_{-66}$ \cr
    $R_{\rm BD}$ (\rj) & 0.776$^{+0.035}_{-0.033}$ & 0.887$^{+0.031}_{-0.031}$ \cr
    $M_{\rm BD}$ (\mj) & 68.0$^{+2.2}_{-2.2}$ & 40.4$^{+1.0}_{-1.0}$ \cr
    $\log{g_{\rm BD}}$ & 5.394$^{+0.030}_{-0.032}$ & 5.105$^{+0.029}_{-0.029}$ \cr
    $\omega$ (deg) & 8.2$^{+9.7}_{-9.1}$ & 94.0$^{+1.6}_{-1.6}$ \cr
    $a$ (au) & 0.0739$^{+0.001}_{-0.001}$ & 0.0647$^{+0.0007}_{-0.0007}$ \cr
    $a/R_\star$ & 37.90$^{+0.91}_{-0.91}$ & 26.76$^{+0.59}_{-0.57}$ \cr
    $R_{\rm BD}/R_\star$ & 0.202$^{+0.004}_{-0.004}$ & 0.175$^{+0.005}_{-0.005}$ \cr
    $M_{\rm BD}/M_\star$ & 0.150$^{+0.003}_{-0.003}$ & 0.072$^{+0.001}_{-0.001}$ \cr
    $i$ (deg) & 89.54$^{+0.24}_{-0.16}$ & 89.59$^{+0.28}_{-0.38}$ \cr
    $t_s$ (\bjdtdb) & 2459604.867$^{+0.024}_{-0.038}$ & 2459632.783$^{+0.052}_{-0.052}$ \cr
    $T_{SB}$ (K) & 1113 & 1049 \cr
    $^{\rm Transit}_{\rm Depth}$ (I) (ppm) & 40000$^{+1800}_{-1900}$ & 35000$^{+2000}_{-2000}$ \cr
    $^{\rm Occultation}_{\rm Depth}$ (K) (ppm) & 500 & 200 \cr
    \hline
    \end{tabular}
    \label{tab: final parameters}
\end{table}

\section{Discussion}
\label{sec: discussion}

\subsection{Population Context}

\begin{figure}[h]
    \centering
    \includegraphics[width=\linewidth]{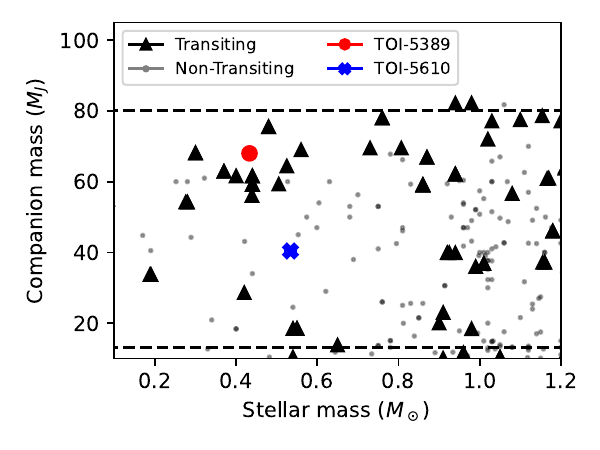}
    \caption{Host mass versus companion mass for known transiting BDs. Black triangles denote transiting systems while gray dots denote non-transiting. The red dot denotes TOI-5389Ab and the blue $\times$ denotes TOI-5610b. The dashed lines mark the upper and lower mass limits for BDs at 80 and 13 \mj, respectively.}
    \label{fig: BD comparison ms vs mp}
\end{figure}



\begin{figure}[h]
    \centering
    \includegraphics[width=\linewidth]{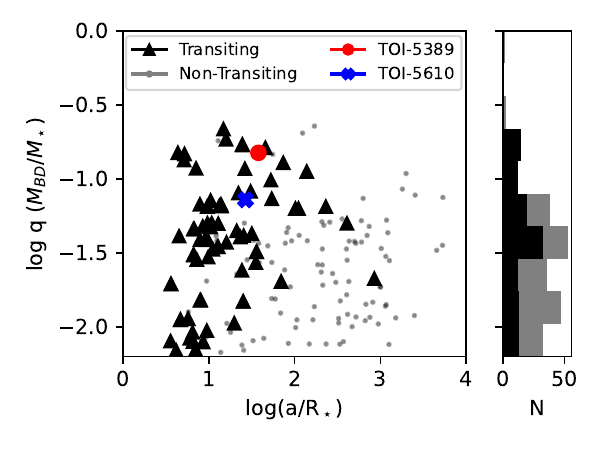}
    \caption{Comparison plot of mass ratio versus semi-major axis. Black triangles denote transiting systems while gray dots denote non-transiting. The red dot denotes TOI-5389Ab and the blue $\times$ denotes TOI-5610b. The histogram on the right shows the distribution in $\log{q}$, transiting systems in black.}
    \label{fig: BD comparison q vs a/AU}
\end{figure}

We constructed a literature comparison sample of BDs, originating largely from exoplanet.eu ($m \sin i < 60 \mj$), \cite{Stevenson2023-driestBD} ($13 \mj < m < 80 \mj$), and \cite{Schmidt2023-bdJune2023} ($13 \mj < m < 80 \mj$), with additions/updates from \cite{Page2024-TOI-1994}, \cite{grieves_bdmd}, \cite{Benni2021-GPX1}, and \cite{Henderson2024-NGTS27Ab}. We limited the sample to BDs with main sequence (MS) host stars and masses $M > 10 \mj$, totaling 393 systems. Of these, 131 are transiting, and the remaining non-transiting systems only yield $m \sin i$, a minimum mass of the BD.

Figure \ref{fig: BD comparison ms vs mp} plots companion mass versus host mass. The dashed lines denote the upper and lower mass limits for BDs at 80 and 13 \mj, respectively. TOI-5389Ab is one of the most massive BDs orbiting an M dwarf and is near the upper mass limit (80 \mj). TOI-5610b lies in the intermediate and less populated (by transits) region around 40 $\mj$. Among transiting BDs, there appears to be a preference towards higher masses, near the upper limit. This should not be a selection effect since radius (and therefore transit depth and transit probability) weakly scales \textit{inversely} with mass for BDs; transiting low-mass brown dwarfs should be \textit{more} likely to be detected. The figure also shows a paucity of low mass BDs around late-type stars. Even low-mass BDs impart a large RV signal on the host star so that they are not likely to be underreported given modern RV sensitivities.  While the comparison sample used here cannot be considered unbiased or complete, these two features considered together could be explained by an evolutionary process in which BDs accrete mass as they undergo inward migration in a gaseous disk, leading to a preponderance of close, high-mass BDs \citep{Bate2002a, Bate2002b, Bate2009, MoeKratter2018, Tokovinin/max2020-migration/accretion}.



Figure \ref{fig: BD comparison q vs a/AU} plots the log of mass ratio versus the log of semi-major axis in units of the host star radius. For non-transiting systems the plotted mass ratio is a lower limit. The paucity of systems with $\log q \gtrsim -1.0$ is a consequence of sample construction, since we do not include stellar companions; furthermore, low-mass hosts comprise only a fraction of the comparison sample. The lack of transiting systems at $a/R_\star >100$ is understood as a consequence of the small separations required to detect a transit via TESS or other systematic transit finding programs. TOI-5389Ab has one of the highest mass ratios for BD-MS systems, and TOI-5610 lies among the upper third. The histogram at right depicts the relative frequencies of $\log q $. There appears to be a preponderance of systems near $q \approx 0.03$ ($\log q \approx -1.5$). Given the probable biases and incompleteness of this comparison sample, we draw no further conclusions from this.

\begin{figure}[h]
    \centering
    \includegraphics[width=\linewidth]{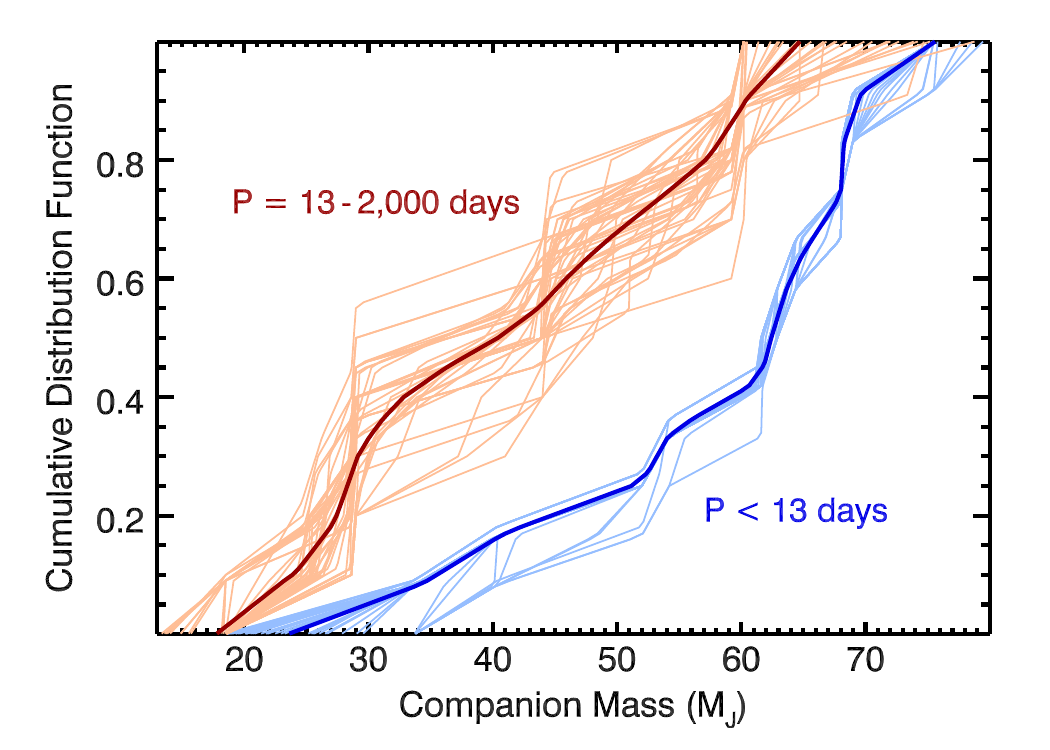}
    \caption{Cumulative distribution function for M-dwarf/BD systems divided into two populations via period.}
    \label{fig: maxplot}
\end{figure}

Compared to the measured occurrence rates of stellar and planetary mass companions, solar-type FGK dwarfs exhibit a dearth of BD companions within $a$~$<$~1~au commonly known as the BD desert \citep{BDdesert, Csizmadia2015}. The mass-ratio distribution of short-period A-dwarf host binaries also exhibits a rapid turnover below $q$~$<$~0.1 \citep{MurphyMoe2019}. Alternatively, BD companions at longer periods are plentiful; the companion mass distribution of solar-type binaries across intermediate separations $a$~=~10\,-\,100~au is relatively uniform above $M_2$ $>$~13~M$_{\rm J}$ \citep{Nielsen2019, Wagner2019}, and exhibits no preference for host mass. Close companions originally fragmentated on large protostellar disk or molecular core scales followed by inward migration as the binary accreted from the surrounding gaseous disk/envelope \citep{Bate2002a, Bate2002b, Bate2009, MoeKratter2018}. Only a small fraction of BD companions can migrate below $a$ $<$ 1 au through the disk without also accreting into the stellar-mass regime \citep{Tokovinin/max2020-migration/accretion}. 

Our population of M-dwarf primaries with BD companions is sufficiently large to probe the nature of the BD desert for low-mass stars. We limit our sample to the 25 systems with $P$ $<$ 2,000 days, $M_\star$ = 0.08\,-\,0.60\,$\msun$, and either $M_{\rm BD}$ = 13\,-\,80~M$_{\rm J}$ or  $M_{\rm BD}$\,sin\,$i$ = 10\,-\,80~M$_{\rm J}$. We divide this sample evenly between the 13 systems with short periods below $P$ $<$ 13 days, including our two objects TOI-5389Ab and TOI-5610b, and the 12 systems across $P$ = 13\,-\,2,000 days. We display the cumulative distribution functions (CDFs) of companion masses for both subsets in Figure~\ref{fig: maxplot}. We adopt the measured companion masses for the eclipsing and astrometric binaries. For the two and six systems with only RV measurements (only $M_{\rm BD}$\,sin\,$i$) in our short and long-period subsets, respectively, we utilize a Monte Carlo technique to simulate random orientations. We synthesize 1,000 CDFs for both samples, where we draw cos\,$i$ = U[0,1] from a uniform distribution for those systems without inclination measurements. We then limit each simulated population to the interval $M_{\rm BD}$ = 13\,-\,80~M$_{\rm J}$, where in most cases we remove the one object with $M_{\rm BD}$\,sin\,$i$ = 10.4 M$_{\rm J}$ and one of the two systems with $M_{\rm BD}$\,sin\,$i$~$\ge$ 54~M$_{\rm J}$. In Figure~\ref{fig: maxplot}, we display the first 50 Monte Carlo simulations and the average CDF for both the short and long-period subsets. 

The short-period subset is weighted toward systematically more massive BD companions compared to the long-period subset, similar to the trend observed for more massive primaries. Only 2/13 = 15\% of the short-period systems have $M_{\rm BD}$ = 13\,-\,40\,M$_{\rm J}$ whereas 6/12 = 50\% of the long-period systems have such low-mass BD companions. For each simulated pair of CDFs, we compute the two-sample Kolmogorov-Smirnov (KS) and Anderson-Darling (AD) statistics. The short and long-period subsets are discrepant with each other at the $p$ = 0.03 (2.2$\sigma$) and $p$ = 0.02 (2.6$\sigma$) probability levels according to the average KS and AD statistics, respectively. This provides the first tentative evidence that M-dwarf primaries exhibit a dearth of $q$~$<$~0.1 ($M_{\rm BD}$ $<$ 40~M$_{\rm J}$) companions at short periods compared to slightly wider systems, consistent with the BD desert observed for AFGK primaries. This suggests that close M-dwarf/BD binaries formed via disk/core fragmentation, inward disk migration, and circumbinary accretion similar to their more massive solar-type counterparts. However, this sample size is still small and combines different detection techniques, and so could have unknown systematics.


\subsection{Age via the Sonora Bobcat Models}
\label{sec: bobcat}

\begin{figure*}
    \centering
    \includegraphics[scale=0.85]{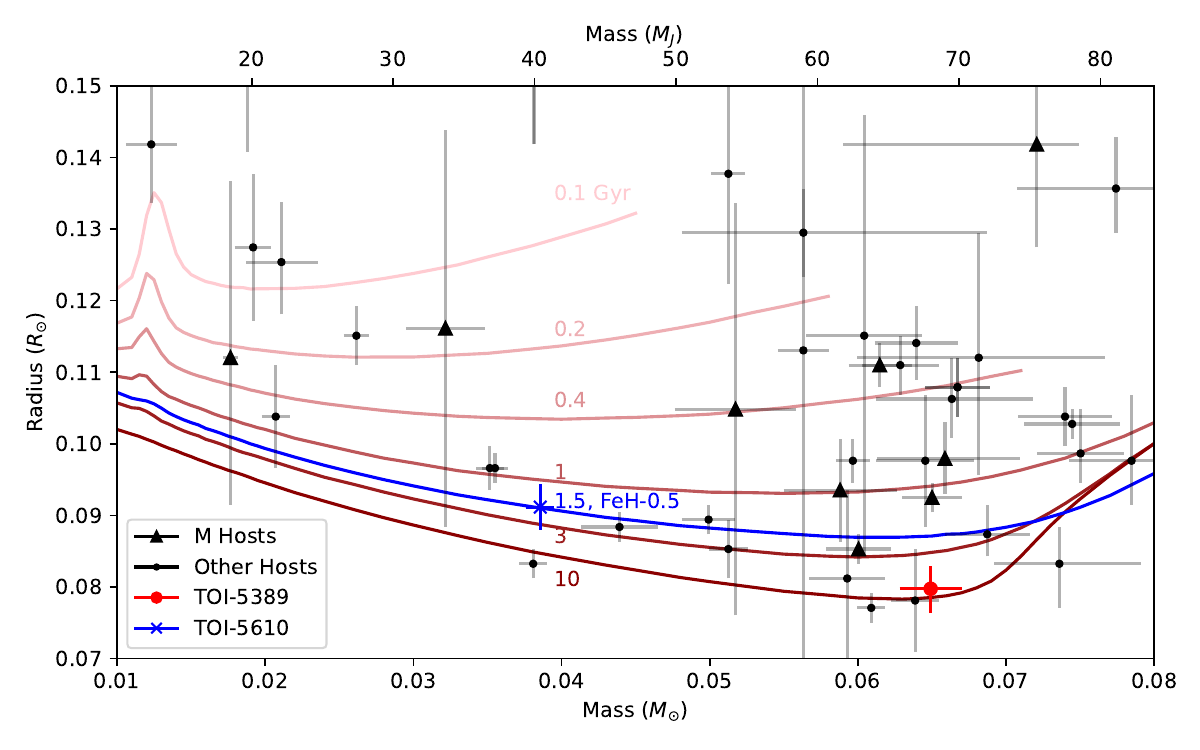}
    \caption{The Sonora-Bobcat models on a plot of companion radius versus mass. The red lines represent isochrones of 0.1, 0.2, 0.4, 1, 3, and 10 Gyr (from light to dark), all at solar metallicity ($\feh=0.0$). The blue line shows the nearest model to TOI-5610b which is 1.5 Gyr at $\feh = -0.5$. Only transiting systems are plotted.}
    \label{fig: bobcat models}
\end{figure*}

Figure \ref{fig: bobcat models} shows the literature sample of transiting BDs and our two targets compared to the Sonora Bobcat BD models \citep{marley_bdatmosphere} on a plot of BD radius versus mass. The red lines represent isochrones of 0.1, 0.2, 0.4, 1, 3, and 10 Gyr (from light to dark), all at solar metallicity (\feh=0.0). The blue line shows the nearest model to TOI-5610b which is 1.5 Gyr at \feh=-0.5. Our best estimates based on these models put TOI-5389Ab at 8 Gyr and TOI-5610b at 1.5 Gyr. These two targets are among the older half of the comparison sample.

\subsection{Predictions of the Secondary Eclipses}

Estimates of the BD temperature from the closest Sonora model corresponding to the adopted BD mass and radius (also consistent with age) indicate an effective temperature of 1113 K for TOI-5389Ab and 1049 K for TOI-5610b. Therefore, internal heating due to gravitational collapse and deuterium fusion dominate the temperature profile of these objects, reflected light would be negligible at these separations, and we use these temperatures to predict secondary depths. The orbital parameters for each brown dwarf listed in Table~\ref{tab: final parameters} predict\footnote{We modeled the lightcurve of each system with the eclipsing binary star package {\tt PHOEBE2 v2.4} \citep{Prsa2016}, using the adopted stellar and BD parameters tabulated above and assuming blackbody spectral energy distributions for the BDs.} 1--2 hour duration secondary eclipses at the reference times $t_s$, despite the appreciable eccentricity of both targets. The estimated BD temperatures result in secondary eclipse depths of $\lesssim 10$ ppm in the Johnson I band for both targets. Predicted eclipses are deeper in Johnson K band, with estimates of 500 ppm for TOI-5389A and 200 ppm for TOI-5610. We searched the TESS photometry for secondary transits and, unsurprisingly, find none at the predicted times, given that the 1\% photometric RMS is far larger than the predicted depths at that bandpass. An accurate prediction for the BD secondary eclipse depth as a function of wavelength would require use of an appropriate BD atmosphere model and is beyond the scope of the nominal secondary transit estimates given here.

\section{Conclusion}
\label{sec: Conclusion}

We have reported the characterization of two brown dwarfs orbiting M dwarfs via four observed transits from TESS and ground-based photometry in conjunction with high-resolution infrared spectroscopy. TOI-5610b has a moderately high eccentricity. TOI-5389Ab has one of the highest companion-to-host mass ratios of transiting BDs ($M_{\rm BD}/M_\star=0.150$), near the hydrogen burning limit. Despite their eccentricities both targets do have potentially observable secondary transits, though shallow. We computed a variety of age estimates. The {\tt EXOFASTv2} SED fit yields $\sim 10$ Gyr for both targets. The WD SED fit to TOI-5389B yields a minimum post-MS age of $\gtrsim 2$ Gyr, which is likely our strongest constraint on the age. The Sonora Bobcat BD models yield 8 and 1.5 Gyr for TOI-5389Ab and TOI-5610b, respectively. Gyrochronology  provide loose limits of $\gtrsim 1$ Gyr. Galactic kinematics for TOI-5610 imply an older star from the thin or thick disk, while kinematic ages are inconclusive for TOI-5389A. A statistical analysis of M-dwarf/BD systems reveals for the first time that those at short orbital periods ($P < 13$ days) exhibit a dearth of $13 \mj < M_{\rm BD} < 40 \mj$ companions ($q$ $<$ 0.1) compared to those at slightly wider separations, similar to the BD desert previously observed for solar-type~primaries.

These two BD characterizations add to the census of brown dwarfs with well-constrained characteristics and provide new data for differentiating formation mechanisms of substellar objects. A compilation of transiting brown dwarf parameters from the literature suggests that high-mass BDs are more abundant than low-mass BDs---a distribution that may find an explanation in terms of BD migration within a gaseous disk during formation.

\section{Acknowledgements}

Based on observations obtained with the Hobby-Eberly Telescope (HET), which is a joint project of the University of Texas at Austin, the Pennsylvania State University, Ludwig-Maximillians-Universitaet Muenchen, and Georg-August Universitaet Goettingen. The HET is named in honor of its principal benefactors, William P. Hobby and Robert E. Eberly.

These results are based on observations obtained with the Habitable-zone Planet Finder Spectrograph on the HET. The HPF team acknowledges support from NSF grants AST-1006676, AST-1126413, AST-1310885, AST-1517592, AST-1310875, ATI 2009889, ATI-2009982, AST-2108512, and the NASA Astrobiology Institute (NNA09DA76A) in the pursuit of precision radial velocities in the NIR. The HPF team also acknowledges support from the Heising-Simons Foundation via grant 2017-0494.

Some of the observations in this paper made use of the NN-EXPLORE Exoplanet and Stellar Speckle Imager (NESSI). NESSI was funded by the NASA Exoplanet Exploration Program and the NASA Ames Research Center. NESSI was built at the Ames Research Center by Steve B. Howell, Nic Scott, Elliott P. Horch, and Emmett Quigley.

This research has made use of the NASA Exoplanet Archive, which is operated by the California Institute of Technology, under contract with the National Aeronautics and Space Administration under the Exoplanet Exploration Program.

The Pan-STARRS1 Surveys (PS1) and the PS1 public science archive have been made possible through contributions by the Institute for Astronomy, the University of Hawaii, the Pan-STARRS Project Office, the Max-Planck Society and its participating institutes, the Max Planck Institute for Astronomy, Heidelberg and the Max Planck Institute for Extraterrestrial Physics, Garching, The Johns Hopkins University, Durham University, the University of Edinburgh, the Queen's University Belfast, the Harvard-Smithsonian Center for Astrophysics, the Las Cumbres Observatory Global Telescope Network Incorporated, the National Central University of Taiwan, the Space Telescope Science Institute, the National Aeronautics and Space Administration under Grant No. NNX08AR22G issued through the Planetary Science Division of the NASA Science Mission Directorate, the National Science Foundation Grant No. AST-1238877, the University of Maryland, Eotvos Lorand University (ELTE), the Los Alamos National Laboratory, and the Gordon and Betty Moore Foundation.

Data presented herein were obtained at the WIYN Observatory from telescope time allocated to NN-EXPLORE through the scientific partnership of the National Aeronautics and Space Administration, the National Science Foundation, and the NSF's National Optical-Infrared Astronomy Research Laboratory.

This work has made use of data from the European Space Agency (ESA) mission
{\it Gaia} (\url{https://www.cosmos.esa.int/gaia}), processed by the {\it Gaia}
Data Processing and Analysis Consortium (DPAC,
\url{https://www.cosmos.esa.int/web/gaia/dpac/consortium}). Funding for the DPAC
has been provided by national institutions, in particular the institutions
participating in the {\it Gaia} Multilateral Agreement.

CIC acknowledges support by NASA Headquarters through an appointment to the NASA Postdoctoral Program at the Goddard Space Flight Center, administered by ORAU through a contract with NASA.

Supported by the National Science Foundation under Grants No. AST-1440341 and AST-2034437 and a collaboration including current partners Caltech, IPAC, the Oskar Klein Center at Stockholm University, the University of Maryland, University of California, Berkeley , the University of Wisconsin at Milwaukee, University of Warwick, Ruhr University, Cornell University, Northwestern University and Drexel University. Operations are conducted by COO, IPAC, and UW.

This work funded by Wyoming NASA Space Grant Consortium, NASA Grant \#80NSSC20M0113.

This work funded by the Wyoming Research Scholars Program at the University of Wyoming.

We thank Jason Eastman and Noah Vowell for helpful conversations regarding the use of EXOFASTv2.

We thank Maxwell Moe for helpful conversations regarding formation mechanisms and biases.

We thank Brock A. Parker for helpful conversations regarding data reduction and the usage of several software. \\

\facilities{HET, HPF, NESSI, RBO, WIYN, TESS, Gaia, NASA Exoplanet Archive, 2MASS, APASS, SDSS, Pan-STARRS, WISE, NESSI, ZTF, ASAS-SN} \\

\software{EXOFASTv2, barycorrpy, HPF-SpecMatch, TESS-Gaia lightcurve, PHEOBE2 v2.4, AstroImageJ} \\

\newpage

\bibliography{sample631}{}
\bibliographystyle{aasjournal}

\end{document}